\documentclass{article}

\usepackage{ifpdf}
\usepackage[bookmarksnumbered,plainpages,hypertex,unicode,hyperindex]{hyperref}

\usepackage{latexsym}
\usepackage{amsfonts}
\usepackage{amssymb}
\usepackage{amsbsy}
\usepackage{amsmath}   
\usepackage{supertabular}

\makeatletter
\begingroup\expandafter\expandafter\expandafter\endgroup
\expandafter\ifx\csname subequations\endcsname\relax
\else
\let\HyOrg@subequations\subequations
\def\subequations{%
\stepcounter{equation}%
\protected@edef\theHparentequation{%
\@ifundefined{theHequation}\theequation\theHequation
}%
\addtocounter{equation}{-1}%
\HyOrg@subequations
\def\theHequation{\theHparentequation\alph{equation}}%
\ignorespaces
}%
\fi
\makeatother

\ifpdf
  \usepackage[dvips]{graphicx}
  \usepackage{color}
  \definecolor{darkgreen}{rgb}{0,.5,0}

      \definecolor{col1}{rgb}{0.125,0.0625,0.125}

\else
  \usepackage[dvips]{graphicx}
\fi

\usepackage[usenames,dvipsnames]{pstricks}
\ifpdf
 \else
  \usepackage{epsfig}
  \usepackage{pst-solides3d}
\fi

\def\mphantom#1{\phantom{#1}}
\def\mphantom#1{#1}
\def\fZ#1{Z}
\def\fZ#1{$#1$}

\def\mphantom#1{\phantom{#1}}
\def\mphantom#1{#1}
\def\fZ#1{Z}
\def\fZ#1{$#1$}

\def\p{\partial}
\def\dfrac#1#2{{\displaystyle\frac{#1}{#2}}}
\def\stTD#1#2{\hbox to 0em{\mathsurround=0em $\stackrel{#1}{\makebox[0pt]{} #2}$\hss} \phantom{#2}}\def\stscript#1#2{\hbox to 0em{\mathsurround=0em ${\scriptstyle\stackrel{#1}{\makebox[0pt]{} #2}}$\hss} \phantom{#2}}\def\stscriptscript#1#2{\hbox to 0em{\mathsurround=0em ${\scriptscriptstyle\stackrel{#1}{\makebox[0pt]{} #2}}$\hss} \phantom{#2}}\def\st#1#2{\mathchoice{\stTD{#1}{#2}}{\stTD{#1}{#2}}{\stscript{#1}{#2}}{\stscriptscript{#1}{#2}}}
\def\comb#1#2#3{{\mathsurround 0pt\hbox to 0pt {\hspace*{#3}\raisebox{#2}{${#1}$}\hss}}}
\def\combs#1#2#3{{\mathsurround 0pt\hbox to 0pt {\hspace*{#3}\raisebox{#2}{${\scriptstyle #1}$}\hss}}}
\def\combss#1#2#3{{\mathsurround 0pt\hbox to 0pt {\hspace*{#3}\raisebox{#2}{${\scriptscriptstyle #1}$}\hss}}}

\def\df{\mathrm{d}}

\def\p{\partial}
\def\dfrac#1#2{{\displaystyle\frac{#1}{#2}}}
\def\stTD#1#2{\hbox to 0em{\mathsurround=0em $\stackrel{#1}{\makebox[0pt]{} #2}$\hss} \phantom{#2}}\def\stscript#1#2{\hbox to 0em{\mathsurround=0em ${\scriptstyle\stackrel{#1}{\makebox[0pt]{} #2}}$\hss} \phantom{#2}}\def\stscriptscript#1#2{\hbox to 0em{\mathsurround=0em ${\scriptscriptstyle\stackrel{#1}{\makebox[0pt]{} #2}}$\hss} \phantom{#2}}\def\st#1#2{\mathchoice{\stTD{#1}{#2}}{\stTD{#1}{#2}}{\stscript{#1}{#2}}{\stscriptscript{#1}{#2}}}
\def\comb#1#2#3{{\mathsurround 0pt\hbox to 0pt {\hspace*{#3}\raisebox{#2}{${#1}$}\hss}}}
\def\combs#1#2#3{{\mathsurround 0pt\hbox to 0pt {\hspace*{#3}\raisebox{#2}{${\scriptstyle #1}$}\hss}}}
\def\combss#1#2#3{{\mathsurround 0pt\hbox to 0pt {\hspace*{#3}\raisebox{#2}{${\scriptscriptstyle #1}$}\hss}}}

\def\EMT{T}
\def\EMTc{\bar{\EMT}}
\def\EMTi{\mathchoice{\combss{\infty}{1.8ex}{0.15ex}\EMT}{\combss{\infty}{1.85ex}{0.15ex}\EMT}{\combss{\infty}{1.25ex}{-0.12ex}\EMT}{\combss{\infty}{1.2ex}{-0.12ex}\EMT}}

\def\Act{\mathcal{A}}

\def\xxx{\chi}
\def\bFem{\mathbf{F}}

\def\x{\mathtt{x}}

\def\eqdef{\doteqdot}

\def\xp#1{\comb{\cdot}{-0.9ex}{0.3ex}{#1}}

\def\EMT{T}
\def\EMTc{\bar{\EMT}}
\def\EMTi{\mathchoice{\combss{\infty}{1.8ex}{0.15ex}\EMT}{\combss{\infty}{1.85ex}{0.15ex}\EMT}{\combss{\infty}{1.25ex}{-0.12ex}\EMT}{\combss{\infty}{1.2ex}{-0.12ex}\EMT}}

\def\Act{\mathcal{A}}

\def\xxx{\chi}
\def\bFem{\mathbf{F}}

\def\x{\mathtt{x}}

\def\eqdef{\doteqdot}

\def\xp#1{\comb{\cdot}{-0.9ex}{0.3ex}{#1}}

\def\Cron#1#2{\mathchoice{\combs{\backslash}{0.1ex}{0.1ex}\delta^{#1}_{#2}}{\combs{\backslash}{0.1ex}{0.1ex}\delta^{#1}_{#2}}{\combss{\backslash}{0ex}{0ex}\delta^{#1}_{#2}}{}{}}

\def\df{\mathrm{d}}

\def\metr{\mathfrak{m}}

\def\metrp{\mathchoice{\comb{-}{-0.9ex}{0ex}\mathfrak{m}}{\comb{-}{-0.9ex}{0ex}\mathfrak{m}}{\combs{-}{-0.75ex}{-0.1ex}\mathfrak{m}}{}{}}

\def\bje{{\mathsurround 0pt\lower.0ex\hbox{${\scriptscriptstyle \mathbf{e}}$}\mspace{-3.4mu}\mathbf{j}}}
\def\je{{\mathsurround 0pt\lower.0ex\hbox{${\scriptscriptstyle e}$}\mspace{-4.5mu}j}}
\def\bjm{{\mathsurround 0pt\lower.0ex\hbox{${\scriptscriptstyle \mathbf{m}}$}\mspace{-5.6mu}\mathbf{j}}}
\def\jm{{\mathsurround 0pt\lower.0ex\hbox{${\scriptscriptstyle m}$}\mspace{-7.0mu}j}}

\def\p{\partial}

\def\eqdef{\doteqdot}
\def\EMT{\mathchoice{\combs{\to}{0.3ex}{-0.2ex}{T}}{\combs{\to}{0.3ex}{-0.2ex}{T}}{\combss{\to}{0.2ex}{-0.2ex}{T}}{\combss{\to}{0.2ex}{-0.2ex}{T}}}
\def\EMTc{\mathchoice{\combs{-}{1.5ex}{0.2ex}{\EMT}}{\combs{-}{1.5ex}{0.2ex}{\EMT}}{\combss{-}{1.1ex}{0.05ex}{\EMT}}{\combss{-}{1.1ex}{0.05ex}{\EMT}}}
\def\EMTi{\mathchoice{\combss{\infty}{1.8ex}{0.15ex}\EMT}{\combss{\infty}{1.85ex}{0.15ex}\EMT}{\combss{\infty}{1.25ex}{-0.12ex}\EMT}{\combss{\infty}{1.2ex}{-0.12ex}\EMT}}

\def\xc{\mathchoice{\comb{\boldsymbol{\cdot}}{-0.1ex}{-0.05ex}x}{\comb{\boldsymbol{\cdot}}{-0.1ex}{-0.05ex}x}{\combs{\boldsymbol{\cdot}}{-0.05ex}{-0.05ex}x}{}{}}

\def\pu{\mathchoice{\comb{\boldsymbol{\cdot}}{-0.1ex}{0.3ex}u}{\comb{\boldsymbol{\cdot}}{-0.1ex}{0.3ex}u}{\combs{\boldsymbol{\cdot}}{-0.05ex}{0.2ex}u}{}{}}

\def\cE{\mathcal{E}}

\def\Act{\mathcal{A}}

\def\Vol{\overline{V}}
\def\dVol{\df\mspace{-2mu}\Vol}

\def\xxx{\chi}

\def\metrEff{\mathchoice{\combs{\sim}{1ex}{0.2ex}\mathfrak{m}}{\combs{\sim}{1ex}{0.2ex}\mathfrak{m}}{\combss{\sim}{0.66ex}{0.05ex}\mathfrak{m}}{}{}}
\def\tmetrEff{\tilde{\vphantom{A}\metrEff}}
\def\metrEffm1{\check{\metrEff}}
\def\KristffEff{\mathchoice{\combs{\sim}{1.5ex}{0.1ex}\Gamma}{\combs{\sim}{1.5ex}{0.1ex}\Gamma}{\combss{\sim}{1.15ex}{0ex}\Gamma}{}{}}

\def\Ngamma{\mathchoice{\combs{\sim}{-0.1ex}{0ex}{\gamma}}{\combs{\sim}{-0,1ex}{0ex}{\gamma}}{\combss{\sim}{-0.15ex}{-0.05ex}{\gamma}}{}{}}
\def\Nvarphi{\mathchoice{\combs{\sim}{0.1ex}{0.05ex}{\varphi}}{\combs{\sim}{0.1ex}{0.05ex}{\varphi}}{\combss{\sim}{0.1ex}{0ex}{\varphi}}{}{}}

\def\ffun{\Phi}

\def\dffun{\ffun}

\def\LF{\mathchoice{\combs{-}{0.3ex}{-0.1ex}\mathcal{L}}{\combs{-}{0.3ex}{-0.1ex}\mathcal{L}}{\combss{-}{0.25ex}{-0.12ex}\mathcal{L}}{}{}}

\def\mass{{\sf m}}

\def\circffun{\st{\circ}{\!\ffun}}
\def\circa{\st{\circ}{a}}
\def\backffun{\tilde{\vphantom{A^\prime}\ffun}}
\def\tcircffun{\st{\circ}{\tilde{\vphantom{A^\prime}\ffun}}}
\def\ps{\mathfrak{s}}

\def\vso{\mathchoice{\comb{\to}{0.4ex}{-0.2ex}V}{\comb{\to}{0.4ex}{-0.2ex}V}{\combs{\to}{0.3ex}{-0.15ex}V}{}{}}

\def\textremb#1{\hspace{-2.1pt}}

\def\textremc#1{}

\def\mifeng#1{#1}
\def\mifrus#1{}
\def\mifengrus#1{}

\begin{document}



\begin{titlepage}
\phantom{.}

\mifeng{%
\begin{center}
{\bf \LARGE Gravitation\\[1ex]  in Unified Scalar Field Theory}\\[4ex]
{\bf\large Alexander A. Chernitskii}\\[2ex]
\small $^1$ \it Department  of Mathematics\\
\small\it St. Petersburg State Chemical Pharmaceutical University\\
\small\it Prof. Popov str. 14, St. Petersburg, 197022, Russia\\[1ex]
\small $^2$\it A.~Friedmann Laboratory for Theoretical Physics\\
\small The Herzen University, 191186 St.-Petersburg, Russia\\[1ex]
\small\rm alexander.chernitskii@pharminnotech.com or chernitskii@friedmannlab.org
\end{center}
\vspace{1ex}
\hrule
\vspace{2ex}
}{}


\vspace{1ex}

\mifrus{\hrule}{}

\vspace{5ex}

\mifengrus{\newpage}{}
\mifeng{%
\begin{center}
{\bf Abstract}
\end{center}
The scalar field of space-time film is considered as unified fundamental field.
The field model under consideration is the space-time generalization of the model for a two-dimensional thin film.
The force and metrical interactions between solitons are considered. These interactions correspond to the electromagnetic and gravitational interactions respectively.
The metrical interaction and its correspondence to the gravitational one are considered in detail.
The practical applications of this approach are briefly discussed.
\vspace{2ex}
}{}


\mifengrus{\vspace{10ex}}{\mifeng{\newpage}{\vspace{15ex}}}

\hrule

\vspace{1ex}

\tableofcontents

\vspace{4ex}

\hrule

\vspace{4ex}

\end{titlepage}

\section{Introduction}
\label{introd}

\subsection{Unified Fundamental Field}
\label{uniff}

The concept of unified {fundamental} field
{has} existed for a  {sufficiently} {long} time.
According to this concept{,}  matter in all  {its} multiformity can be considered as a solution of some nonlinear field model.

In particular, all elementary particles must be represented by soliton solutions of the model.
 {In this case} their interactions are the consequence of the field model nonlinearity.

The nonlinearity {violates}  the superposition property  {for} solutions  {of} the appropriate linearized model. According to this property the sum of solutions is also solution.
The  {violation} of the superposition property is
interpreted as the interaction of  particles.

It is reasonable that the interaction of big  {material} objects containing many particles is explained by the same way.

{Attempts to propose models of the unified field theory have been made by many well-known researchers.}
It is know{n} that the creator of general relativity theory A. Einstein
was adherent of this concept.
{He tried to create a unified field theory for the rest of his life.}

\subsection{Attempts  {to Create a} Unified Field Theory}
\label{attcreation}

Electromagnetic field satisfied nonlinear equations was considered as unified field.
First of all we must notice two known nonlinear electrodynamics model, namely, Mie~\cite{Mie1912} and Born---Infeld models~\cite{BornInfeld1934a}.

Einstein considered,
in particular, the field of metrical tensor for curved space-time as unified field~\cite{Einstein1930aE}.
In this connection
nonsymmetrical metric was also considered~\cite{Einstein1954aE}.

Heisenberg  {tried to build}
a unified theory of spinor field~\cite{Heisenberg1966e}.

All these attempts  {were} not {completely} successful.
In particular,
{within the framework of these theories, no solution has been obtained that corresponds with a sufficient degree of realism to any elementary particle.}
It should be noted that this problem is
extremely difficult mathematically.

\subsection{{The} Question  {as to the} Tensor Rank of the Field}
\label{qtensranc}

It is evident that the question  {as to the} tensor rank of a unified field is highly~important.

In the above-mentioned attempts to construct a unified field theory,  the choice of the tensor rank of the field was based on certain correspondence considerations.

For example, the nonlinear electrodynamics models  {generalized} the linear electrodynamics which was successful in certain limits.

The  {variants} of Einstein's unified theories  {generalized} his {own} gravitational theory which was successful in certain limits.

Heisenberg's  theory of a unified spinor field  was based on Dirac's  linear theory which was successful in certain limits.

\subsection{{The} Force and Metrical Interactions of Particles-Solitons}
\label{forceandmetr}

As a natural  criterion {for the acceptability} of the unified field  {model,
we should consider the possibility in its framework for description
of the two long-range interactions of material objects, namely electromagnetism and gravitation.}

{A number of the author's works were devoted to the}
unification  {of} electromagnetism and gravitation
in the framework of Born---Infeld nonlinear electrodynamics  (see, for example,
\cite{Chernitskii2004a} -- \cite{Chernitskii1999})).

 {In this case,} the electromagnetic interaction of particles-solitons is a consequence of {the} integral conservation law for energy-momentum. This type of interaction is called the force one.

The description of  {gravity} in the framework of nonlinear electrodynamics is based on the effect of induced space-time curvature by  {a weak} field of distant particles-solitons
 {at the location of the test} particle.
{This type of interaction is called the metrical one.}

\subsection{Scalar Field}
\label{scalfield}

Force and metric interactions of particles-solitons are  {inherent in} all nonlinear field models  {that} are invariant under shifts and  {rotations} in four-dimensional space-time.

Thus we can consider a  {simpler} model  {of a} scalar unified field,
{since it also allows us to describe electromagnetism and gravity}~\cite{Chernitskii2016b}.

In particular, the tensor character of {the} electromagnetic field is
 {due to}
the determination
{of the integral force through the integral over a closed surface surrounding the test particle under}
the force interaction
\cite{Chernitskii2017a}.

It should be  {emphasized} that the electromagnetic field of antisymmetric
tensor {of the second rank} appear{s} at  {a certain} point when
{a test} particle  {is placed there}.
{In the absence of such a particle, only a configuration of the scalar field is present near this point, but it generates} the electromagnetic interaction with any charged particle.

\section{{The} Unified Field}
\label{worldva}

\subsection{Space-Time Film}
\label{stfilm}

\begin{subequations}\label{402655631}

We consider the following generally covariant world volume action and the appropriate variational principle~\cite{Chernitskii2018a}:
\begin{equation}
\label{35135655}
\mphantom{
\Act =\int\limits_{\Vol}\!\sqrt{|\mathfrak{M}|}\;(\mathrm{d}x)^{4}
= \int\limits_{\Vol}\LF\;\dVol
\;,
\qquad
\delta \Act = 0\;,
}
\end{equation}
where \fZ{\mathfrak{M} \eqdef \det(\mathfrak{M}_{\mu\nu})},
\fZ{\left(\df x\right)^{4} \eqdef \df x^{0}\df x^{1}\df x^{2}\df x^{3}},
\fZ{\Vol} is space-time volume, \fZ{\dVol \eqdef \sqrt{|\metr|}\;\left(\df x\right)^{4}} is four-dimensional volume element, {\fZ{\metr\eqdef\det\left(\metr_{\mu\nu}\right)},}

\begin{equation}
\label{381936901}
\mphantom{
\mathfrak{M}_{\mu\nu} = \metr_{\mu\nu} + \xxx^2\,\frac{\p \ffun}{\p x^{\mu}}\,\frac{\p \ffun}{\p x^{\nu}}
\;,
\qquad
\LF \eqdef
\sqrt{\left|1 + \xxx^{2}\,\metr^{\mu\nu}\,\frac{\p \ffun}{\p x^{\mu}}\,\frac{\p \ffun}{\p x^{\nu}}\right| }
}
\end{equation}
\noindent
\fZ{\metr_{\mu\nu}} are components of metric tensor for flat four-dimensional space-time,
\fZ{\ffun} is scalar real field function,
\fZ{\xxx} is dimensional constant.
The Greek indices take values \fZ{\{0,1,2,3\}}.
The tensor \fZ{\mathfrak{M}_{\mu\nu}}  can be called  {the} world tensor.

\end{subequations}

The model (\ref{402655631}) can be considered as a relativistic generalization
of the appropriate expression for the mathematical model of two-dimensional minimal thin film
in the tree-dimensional space of our everyday experience.

\subsection{Energy-Momentum Density Tensor}
\label{emdten}

Customary method gives the following canonical energy-momentum density tensor of the model in
Cartesian coordinates
\begin{equation}
\label{442613621}
\mphantom{
\EMTc^{\mu\nu} = \frac{1}{4\pi}\left(\frac{\dffun^{\mu}\,\dffun^{\nu}}{\LF}-
\frac{\metrp^{\mu\nu}}{\xxx^2}
\,
\LF
\right)
\;,\quad
\ffun^{\alpha} \eqdef \metrp^{\alpha\beta}\,\frac{\p \ffun}{\p x^{\beta}}
\;.
}
\end{equation}
\noindent
where \fZ{\metrp^{\mu\nu}} is the constant diagonal metrical tensor for flat space-time with signature \fZ{\{+,-,-,-\}} or \fZ{\{-,+,+,+\}}.
As we see, the canonical tensor is symmetrical.

To obtain finite integral characteristics of solutions in infinite space-time we introduce {the} regularized energy-momentum density tensor with the following formula:
\begin{equation}
\label{809745461}
\mphantom{
\EMT^{\mu\nu} = \EMTc^{\mu\nu} - \EMTi^{\mu\nu}
\;.
}
\end{equation}
\noindent
where \fZ{\EMTi^{\mu\nu}} is {a} regularizing symmetrical energy-momentum density tensor.
Here we will use {the} constant regularizing tensor
\begin{equation}
\label{431999421}
\mphantom{
\EMTi^{\mu\nu} = -\frac{1}{4\pi\,\xxx^2}\,\metrp^{\mu\nu}
\;.
}
\end{equation}

{%
In general case we can take a symmetrical tensor satisfying the differential conservation law as the
regularizing tensor. A special choice of this tensor can provide the convergence of energy integral for
a certain class of solutions.}

\subsection{Equation of Space-Time Film}
\label{eqsptfilm}

The variational principle (\ref{402655631}) gives the following model field equation in Cartesian coordinates~\cite{Chernitskii2018a}:
\begin{equation}
\label{371394071}
\mphantom{
\left(\metrp^{\mu\nu}\left(1 + \xxx^{2}\,\metrp_{\sigma\rho}\,\ffun^{\sigma}\,\ffun^{\rho}\right) - \xxx^{2}\,\dffun^{\mu}\,\dffun^{\nu}\right)
\frac{\p^{2}\,\ffun}{\p x^{\mu}\,\p x^{\nu}} = 0
\;,
}
\end{equation}

\begin{subequations}\label{45213523}

This field equation can be written in the following  {remarkable} form~\cite{Chernitskii2016b}:
\begin{align}
\label{402146881}
\mphantom{
\metrEff^{\mu\nu} \frac{\p^{2}\,\ffun}{\p x^{\mu}\,\p x^{\nu}}
}
&= 0
\;,\\
\label{402146882}
\metrEff^{\mu\nu} &
\mphantom{
\eqdef -4\pi\,\chi^2\,\EMTc^{\mu\nu}
}
\;.
\end{align}
\noindent
where \fZ{\EMTc^{\mu\nu}} is {the} canonical energy-momentum density tensor (\ref{442613621}).
Here we introduce the effective metric \fZ{\metrEff^{\mu\nu}} which will be considered below.

\end{subequations}

Equation (\ref{45213523}) transforms to ordinary linear wave equation with \fZ{\xxx = 0}:
\begin{equation}
\label{414686341}
\mphantom{
\metrp^{\mu\nu}
\frac{\p^{2}\,\ffun}{\p x^{\mu}\,\p x^{\nu}} = 0
\;.
}
\end{equation}

\subsection{Effective Metric and Curved Space-Time}
\label{effmetr}

The  characteristic equation {of the model} has the following form which is obtained directly from (\ref{402146881}):
\begin{equation}
\label{582613611}
\mphantom{
\metrEff^{\mu\nu}\,\tilde{k}_\mu\,\tilde{k}_\nu = 0
\;,\quad\quad\quad
\tilde{k}_\mu {}\eqdef{} \frac{\partial {\cal S}}{\partial x^\mu}
\;,
}
\end{equation}
\noindent
Equation \fZ{{\cal S}= 0} gives a three-dimensional characteristic hypersurface of the field model in four-dimensional space-time.

As is known, the problem of propagation  {of a} weak high-frequency quasi-plane wave with {a} given background
field in {a} nonlinear model gives the dispersion relation for {the components of the} wave four-vector
 {that} coincides
with {the} characteristic equation~\cite{Chernitskii2012be}.
{The} dispersion relation or a dependence between {the} frequency of the wave and its {three-dimensional} wave vector defines
the wave front propagation.

Thus the components \fZ{\tilde{k}_\mu} in Equation~(\ref{582613611}) can {also} be considered as {the components of the} wave four-vector
of {a} high-frequency quasi-plane wave propagating in {an} effective curved space-time with {the} metric \fZ{\metrEff^{\mu\nu}}.

Equation (\ref{582613611}) sets  the {motion of such} wave packets
{that, in the absence of a background field, would move at}
 the speed of
light. Thus this equation  {determines} the
 {trajectories} of
massless particles.
{For the} trajectory of a massive particle{, the considered approach gives the equation of a geodesic line in an effective curved space-time with the metric \fZ{\metrEff^{\mu\nu}}}.
This theme is considered below in Section
\ref{fundint}.

\section{Solitons-Particles}
\label{solparts}

\subsection{Solitons are Material Particles}
\label{solarematpart}

Solitons are  {spatially} localized solutions of nonlinear field models. Solitons-particles are
soliton solutions of the unified field model  {corresponding} to elementary particles.

Another name for soliton is solitary wave. But linear field models
 can {also} have the solutions in the form of solitary waves.
 {Because} of linearity of the models these solutions can have an arbitrarily small amplitude. Thus we shall call these solitary waves the weak solitons.

Not long ago, a class of exact soliton solutions propagating with the speed of light was obtained
for the model of space-time film~\cite{Chernitskii2018a}. A subclass of these solutions has properties which allow to correlate
it with photons. Other {subclasses of these} solutions
{may} relate to  {various types} of neutrino.

At the present time, among exact soliton solutions for massive particles we know
the simplest spherically symmetrical soliton in intrinsic coordinate system or spheron~\cite{Chernitskii2017a}.
It  {has} a rest energy and electrical charge but do{es} not have angular momentum or spin.

There is a defined progress in the finding of approximate soliton solutions  {of the} toroidal configuration
 {containing an} oscillating part~\cite{Chernitskii2020a}.
These solutions can  {also have} a spin in addition to  {the} rest mass and charge.

\subsection{Oscillating Parts of Solitons-Particles}
\label{oscparts}

First of all the  {existence} of oscillating parts  {in} solitons-particles
{is} necessary for the description
of observable wave properties for elementary particles.

But in the framework of the unified field theory under consideration the oscillating parts
are necessary for the description of real gravitation. We shall discuss this question below.

When we move away from the localization region of a soliton-particle{,} the field of space-time film satisf{ies} approximately the linear
wave Equation~(\ref{414686341}).

\begin{subequations}\label{647852471}

Elementary oscillat{ing} solutions of the linear wave equation in the spherical coordinate system decrease in amplitude as \fZ{r^{-1}}.
Let us take the following simple solution as example:
\begin{equation}
\label{413418791}
\mphantom{
\circffun = \frac{\circa}{\xp{r}}\sin(\xp{\omega}\,\xp{r})\,\sin(\xp{\omega}\,\xp{x}^{0})
\;,
}
\end{equation}
\noindent
where the point under symbols denotes the belonging to the intrinsic coordinate system.

According to the definition given in the previous subsection{,} this solution is a weak soliton.

The asymptotic form such as we have in (\ref{413418791})  {leads to} the divergence of energy integral at infinity.

 {However,} the investigation of toroidal configurations~\cite{Chernitskii2020a} shows that {there can be}
 solitons with oscillat{ing} part
without the asymptotic form of type (\ref{413418791})  {whose energy is finite}.
{But} we can suppose in this case that the wave mode of type (\ref{413418791})  {will} appear for interacting solitons.

Thus we assume that a soliton-particle in {the} intrinsic coordinate system has both static and
oscillat{ing} parts.
{This representation is appropriate to usual decomposition of time-periodic solution in temporal Fourier series.
Of course, the static and oscillating parts of a soliton-particle are not separate soliton solutions.}

The oscillating part is a standing wave having perhaps sufficiently complicated configuration.
Using a space-time rotation we can obtain a moving soliton from the rest one.
In this case the standing wave transforms to moving one.
The moving soliton is obtained with {the} help of the following
substitution for the intrinsic coordinates of the soliton:
\begin{equation}
\label{617928341}
\mphantom{
\xp{x}^{\mu} = L^{\mu}_{.\nu}\,x^{\nu}
\;,
}
\end{equation}
\noindent
where \fZ{L^{\mu}_{.\nu}\,L^{\rho\nu} = L^{.\mu}_{\nu}\,L^{\nu\rho} = \metrp^{\mu\rho}}, \fZ{L^{\mu}_{.\nu}} is a matrix of space-time rotation,
in particular, Lorentz transformations, \fZ{\{x^{\nu}\}} is the coordinate system in which the soliton is moving.

\end{subequations}

We have the following dispersion relation for the wave {four-}vector components of this moving wave:
\begin{equation}
\label{66874556}
\mphantom{
\left|\metrp^{\mu\nu}\,k_{\mu}\,k_{\nu}\right| = \xp{\omega}^2
\;,\quad{\omega\eqdef -k_0\;,}
}
\end{equation}
\noindent
where \fZ{\xp{\omega}} is  {the angular} frequency of the standing wave in intrinsic coordinate system of the soliton,
\fZ{\{\omega,k_i\}} are {the frequency and} wave vector components of the traveling wave.

 {As can be verified {\cite{Chernitskii2012be}}, the} group velocity {of the traveling wave} corresponding
{to} the dispersion relation (\ref{66874556})
coincides with the velocity parameter {\fZ{\vso}} of the space-time rotation~{\mbox{(\ref{617928341})}}.
The frequency \fZ{\xp{\omega}} is called a rest frequency of the soliton.

\section{Gravitation}
\label{fundint}

\subsection{Gravitation as {the} Metrical Interaction of Solitons}
\label{gravasmetrin}

The effect of induced gravitation in nonlinear electrodynamics
{is the subject of a sufficient number of}
works  {by the} author
\cite{Chernitskii2004a} -- \cite{Chernitskii2016b}, \cite{Chernitskii2011a} -- \cite{Chernitskii2009a},
including  {an} article in {the} encyclopedia~\cite{Chernitskii2004a} and {a}
monograph~\cite{Chernitskii2012be}.

Now this analysis is applied to the space-time film model with insignificant modification.

Let us consider the problem  {of} propagation of a small amplitude wave  {with}
a rest frequency against a background given field of distant solitons-particles.

For distinctness we take the simple solution of the linear wave Equation~(\ref{413418791}). This solution has the
form of {a} standing wave in {the} intrinsic coordinate system.
It has a finite amplitude  {at the} coordinate origin \fZ{\circa} which can be
taken
sufficiently small.

\begin{subequations}\label{294337661}
Let us consider the sum of {the} background field
\fZ{\backffun}
generated by distant solitons and a moving {fast-oscillating} weak soliton \fZ{\tcircffun} of type (\ref{647852471})
but with a  {slow time-dependent velocity} parameter {\fZ{\vso}} and
{a} small
{constant}
amplitude  {\fZ{\circa}}:
\begin{align}
\label{323630831}
&\mphantom{\ffun = \backffun + \tcircffun}
\;,\\
\label{323630832}
&\mphantom{\left|\omega\right| \gg \Bigl.\max\Bigl|\frac{\p\backffun}{\p x^{\mu}}\Bigr|\Bigr/\!\max\bigl|\backffun\bigr|
\;,\quad \Bigl|\frac{\df\vso}{\df x^{0}}\Bigr| \ll \left|\omega\right|
\;,\quad \circa \ll \max\bigl|\backffun\bigr|}
\;.
\end{align}
\end{subequations}

{Here we consider the weak soliton with a constant amplitude \fZ{\circa} as some approximation for a soliton-particle whose amplitude is defined by an exact solution.
If there is an additional field of remote solitons \fZ{\backffun}, the soliton solution is modified,
and its amplitude could also be changed.
However, it is natural to assume
that the soliton-particle has a maximum amplitude which {is} significantly greater than the field of distant solitons-particles.
Then the weak field of distant particles will not significantly affect the amplitude of the considered soliton-particle.
Thus if we consider the weak soliton instead of the soliton-particle, we actually investigate the influence of the distant solitons field \fZ{\backffun} to the part of soliton-particle which is sufficiently far from its center and has a small amplitude.
The movement of this weak part should direct the entire soliton-particle, since it is a modified exact solution.}

We substitute this sum {(\ref{323630831})} in the equation of space-time film (\ref{45213523}).
{According to the made assumptions (\ref{323630832}), we keep only the terms with the second derivatives of the weak soliton field \fZ{\tcircffun} in the obtained equation.}

Here we suppose an averaging of
 {the effective metric \fZ{\metrEff^{\mu\nu}} (\ref{402146882}) with}
the background field \fZ{\backffun} over a defined space-time localization region for the weak soliton \fZ{\tcircffun}.
{The averaged components of the effective metric are the constant components of the decomposition for the effective metric components in four-dimensional Fourier series in this soliton localization region.}

Then we obtain the following equation:
\begin{equation}
\label{387513291}
\mphantom{
\tmetrEff^{\mu\nu}
\frac{\p^{2}\,\tcircffun}{\p x^{\mu}\,\p x^{\nu}} = 0
\;.
}
\end{equation}
\noindent
where the averaged effective metric \fZ{\tmetrEff^{\mu\nu}}
depends on the derivatives of the background field:
\fZ{\tmetrEff^{\mu\nu} = \tmetrEff^{\mu\nu} (\backffun)}.
{Here below we use the designation
\fZ{\metrEff^{\mu\nu}}
for the averaged effective metric
\fZ{\tilde{\metrEff}^{\mu\nu}}
in an effort to simplify the designations.}

We consider that the averaged {background} effective metric \fZ{\metrEff^{\mu\nu}} is almost  {constant}
in  {the} space-time  {localization region of} the weak soliton \fZ{\tcircffun}.
In this region, let us find a coordinate transformation which reduces the Equation~(\ref{387513291}) to ordinary wave one (\ref{414686341}).

Let we have the following relations for {the} coordinate differentials:
\begin{equation}
\label{Def:TransG}
\mphantom{
\df \xp{x}^\mu = \hat{X}^\mu_{.\nu}\,\df x^\nu
\;, \quad
\df x^\mu = \check{X}^\mu_{.\nu}\,\df \xp{x}^\nu
\;,
}
\end{equation}
\noindent
where the matrix \fZ{\hat{X}^\mu_{.\nu}} and \fZ{\check{X}^\mu_{.\nu}} are mutually inverse and satisfy the relations

\begin{equation}
\label{468596101}
\mphantom{
\metrEff^{\mu\nu}\,\hat{X}^\sigma_{.\mu}\,\hat{X}^\rho_{.\nu} = \metrp^{\sigma\rho}
\;,\quad
\metrEff^{\mu\rho} = \check{X}^\mu_{.\zeta}\,
\check{X}^\rho_{.\xi}\,\metrp^{\zeta\xi}
\;.
}
\end{equation}

Then the solutions of Equation~(\ref{387513291}) in the limited  {region} under consideration has the form
\begin{equation}
\label{Sol:tAp}
\mphantom{
\tcircffun =
\circffun \bigl(\xp{x}^\sigma (x)\bigr)
\;.
}
\end{equation}
\noindent
where ~\fZ{\circffun \bigl(\xp{x}^\sigma\bigr)} is a solution of the linear wave Equation~(\ref{414686341}), in particular, the weak soliton~\mbox{(\ref{413418791})}.
Here the functions \fZ{\xp{x}^\sigma (x^{\delta})} are defined by the transformation (\ref{Def:TransG}).

The averaged background
effective metric \fZ{\metrEff^{\mu\nu}} calculated in the expanded four-dimensional space defines a Riemann
space which is not generally flat.
Also, in general, it is not possible to
{find}
the coordinate transformation \fZ{\xp{x}^\sigma = \xp{x}^\sigma (x^{\delta})} satisfying (\ref{Def:TransG}) and (\ref{468596101})
everywhere.

\begin{subequations}\label{561463141}
But we can write this transformation approximately in the following form:
\begin{align}
\label{Def:yG}
\xp{x}^i &=
\mphantom{\hat{X}^i_{.j} (x^0)\,\left(x^j {}-{} \xc^j(x^0) \right)}
\;,\\
\label{Def:bark}
\xp{x}^0 &=
\mphantom{\xp{x}^0 (x^\nu) = \int \bar{k}_\mu\,\df x^\mu
\;,\quad
\bar{k}_\mu (x^\nu) \eqdef \hat{X}^0_{.\mu} (x^\nu)}
\;.
\end{align}
\noindent
where \fZ{\xc^j(x^0)} is a position of energy center for the weak soliton at the time \fZ{x^0},
\fZ{\bar{k}} is called a normalized wave vector such that
\begin{align}
\label{Rel:dkdx}
\frac{\p \bar{k}_\mu}{\p x^\nu} &=
\mphantom{\frac{\p \bar{k}_\nu}{\p x^\mu}}
\;.
\end{align}
\end{subequations}

Taking into account the transformation (\ref{561463141}) we obtain that \fZ{\xp{\omega}\,\xp{x}^0} is a phase of the wave \fZ{\tcircffun}.
Then we have the following its wave vector components:
\begin{equation}
\label{519646101}
\mphantom{
k_\mu {}={} \xp{\omega}\,\bar{k}_\mu
\;.
}
\end{equation}

{It should be emphasized that
the substitution of type (\ref{Def:bark}) for \fZ{\xp{x}^0} to the solution (\ref{413418791}) is the known approach to solving
the problem of wave propagation in an inhomogeneous medium
\cite{Whitham1974}.
In this approach, the wave phase is considered as an unknown coordinate function, and the substitution of the modified wave solution
into the equation gives a dispersion relation for this wave.}

Using (\ref{468596101}) and (\ref{Def:bark}), we have the following dispersion relation for the  {weak} soliton:
\begin{align}
\label{Rel:Disper}
\mphantom{\metrEff^{\mu\nu}\,\bar{k}_\mu\,\bar{k}_\nu }
&= \metrp^{00}
\;,
\end{align}
\noindent
where \fZ{\metrEff^{\mu\nu} {}={} \metrEff^{\mu\nu}(x^\rho)},
\fZ{\bar{k}_\mu {}={} \bar{k}_\mu (x^\rho)}.

\begin{subequations}\label{742618171}
Now let us obtain a trajectory of energy center for the weak soliton \fZ{\xc^j= \xc^j(x^0)}.
We use the intrinsic time \fZ{\xp{x}^0 (x^\mu)} (\ref{Def:bark}) at the point \fZ{\xc^\mu} as a parameter of movement \fZ{\ps}.
Taking also \fZ{\xc^0 = x^0} we have
\begin{align}
\label{738309291}
\ps &
\mphantom{\eqdef \xp{x}^0(\xc^\mu) = \ps (x^{0})}
\;,\\
\label{Def:U}
\pu^\mu &
\mphantom{\eqdef \dfrac{\df \xc^\mu}{\df \ps}
= \check{X}^\mu_{.0}(\xc^\nu)}
\;.
\end{align}
\noindent
where \fZ{\pu^\mu} is  {the} four velocity. We have the last relation in (\ref{Def:U}) in accordance to (\ref{Def:TransG}) and~\mbox{(\ref{738309291})}.
\end{subequations}

\begin{subequations}\label{Rel:kU}
The definitions \fZ{\pu^{\mu}} (\ref{Def:U}) and \fZ{\bar{k}_\nu} (\ref{Def:bark}) with (\ref{468596101})
gives also the following relations at the point \fZ{\{\xc^\mu\}}:
\begin{align}
\label{543881161}
&\mphantom{\bar{k}_\mu\,\pu^\mu {}={} 1}
\;,\\
\label{55183142}
&\mphantom{\pu^\mu {}={} \metrp^{00}\,\metrEff^{\mu\nu}\,\bar{k}_\nu}
\;.
\end{align}
\end{subequations}

Let us introduce  {the} inverse tensor \fZ{\metrEffm1_{\mu\nu}} to the tensor \fZ{\metrEff^{\mu\nu}}:
\begin{align}
\label{Def:invg}
\mphantom{\metrEffm1_{\mu\nu}\,\metrEff^{\nu\rho} }
&= \Cron{\rho}{\mu}
\;.
\end{align}

We have from (\ref{Rel:kU}) and (\ref{Def:invg}) the following relation at the point \fZ{\{\xc^\mu\}}:
\begin{align}
\mphantom{\metrEffm1_{\mu\nu}\,\pu^\mu\,\pu^\nu }
&= \metrp^{00}
\;.
\end{align}
\noindent
{This leads to the well-}%
known expression in general relativity theory:
\begin{align}
\df \ps^2 &= \mphantom{\dfrac{1}{\metrp^{00}}\,
\metrEffm1_{\mu\nu}\,\df\xc^\mu\,\df\xc^\nu
{\,}={\,}
\left|\metrEffm1_{\mu\nu}\,\df\xc^\mu\,\df\xc^\nu\right|}
\;.
\end{align}

We have also from (\ref{Rel:kU}) the following relation at the point \fZ{\xc^\mu}
\begin{align}
\label{Expr:kgU}
\bar{k}_\mu &= \mphantom{\dfrac{1}{\metrp^{00}}\,\metrEffm1_{\mu\nu}\,\pu^\nu}
\end{align}

 One must note that although \fZ{\pu^\mu} and \fZ{\bar{k}_\mu} are defined as
 components of second-order matrix,
they are actually four-vectors.

Indeed, if we consider the problem in another coordinate system \fZ{\{x^{\prime\mu}\}} then we have
\begin{equation}
\mphantom{
\df\ps\eqdef \df \xp{x}^0 {}={} \hat{X}^0_{.\mu}\,\df x^\mu {}={}
\hat{X}^{\prime 0}_{.\mu}\,\df x^{\prime\mu}
\;.
}
\end{equation}
\noindent
This means that the intrinsic time \fZ{\xp{x}^0} is not changed for such transformations or it behaves as scalar.
Thus \fZ{\bar{k}_\mu} is a four-vector according to the definition (\ref{Def:bark}).

{As noted above, \fZ{\xp{\omega}\,\xp{x}^{0}} is the phase of the wave which is, of course, relativistic invariant. Then the function \fZ{\xp{x}^{0} (x^{\mu})}
can be called the normalized phase. Thus we can write the following definition for the normalized wave vector which is in agreement with (\ref{Def:bark}) and \mbox{(\ref{Rel:dkdx})}:}
{%
\begin{equation}
\label{778527301}
\mphantom{ \bar{k}_\mu \eqdef \frac{\p \xp{x}^{0}}{\p x^{\mu}}}
\;.
\end{equation}}

Then, according to (\ref{Def:yG}), the three-dimensional coordinate system \fZ{\{\xp{x}^i\}} is moving coupled with the localization region of the weak soliton.
Thus we have {\fZ{\xp{x}^i {}={} 0}} and \fZ{\df \xp{x}^i {}={} 0} at the point \fZ{\{\xc^\mu\}} and
\begin{equation}
\label{559894451}
\mphantom{
\df \xc^\mu = \check{X}^\mu_{.0}(\xc^\nu)\,\df \xp{x}^0
\;,\quad
\df \xc^{\prime\mu} =
\check{X}^{\prime\mu}_{.0}(\xc^{\prime\nu})\,\df \xp{x}^0
\;.
}
\end{equation}
\noindent
Therefore \fZ{\pu^\mu} are a four-vector according to definition (\ref{Def:U}).

Now let us obtain the trajectory equation for \fZ{\xc^\mu (\ps)}.
Differentiation of dispersion relation (\ref{Rel:Disper}) with respect to certain coordinate \fZ{x^\rho} with consideration of relations (\ref{Rel:dkdx}) and~\mbox{(\ref{Rel:kU})}
gives the following equation at the point \fZ{\xc^\mu}:
\begin{align}
&\mphantom{\dfrac{\p \metrEff^{\mu\nu}}{\p x^\rho}\,\bar{k}_\mu\,\bar{k}_\nu
{}+{} 2\,\metrp^{00}\,\pu^\nu\,\dfrac{\p \bar{k}_\rho}{\p x^\nu}
{}={} 0}
\mphantom{\quad\quad\Longrightarrow\quad\quad
\dfrac{\p \metrEff^{\mu\nu}}{\p x^\rho}\,\bar{k}_\mu\,\bar{k}_\nu
{}+{} 2\,\metrp^{00}\,\dfrac{\df \bar{k}_\rho}{\df \ps}
{}={} 0}
\;.\qquad
\label{Eq:Geodtr}
\end{align}

\begin{subequations}\label{564676361}
Substituting (\ref{Expr:kgU}) {in}to (\ref{Eq:Geodtr}) and using
(\ref{Def:invg}) we obtain the trajectory equation in the following form:
\begin{align}
\label{56548621}
\mphantom{
\dfrac{\df \pu^\mu}{\df \ps} {}+{} \KristffEff^\mu_{\nu\rho}\,\pu^\nu\,\pu^\rho}
&= 0
\;,
\end{align}
\noindent
where
\begin{align}
\label{56555770}
\KristffEff^\mu_{\nu\rho} &\mphantom{\eqdef
\dfrac{1}{2}\,\metrEff^{\mu\delta}\left(
\dfrac{\p\metrEffm1_{\delta\nu}}{\p x^\rho} {}+{}
\dfrac{\p\metrEffm1_{\delta\rho}}{\p x^\nu} {}-{}
\dfrac{\p\metrEffm1_{\nu\rho}}{\p x^\delta}
\right)}
\;.
\end{align}
\end{subequations}

{As can be seen,} this equation is the geodesic line one for the introduced effective Riemann space with {the} metric \fZ{\metrEff^{\mu\nu}}.

Thus a weak soliton with rest frequency {and small constant amplitude} under the influence of distant solitons behaves as massive particle in gravitational field.

We {can} assume that the obtained Equation~(\ref{564676361}) describes also the movement of a soliton-particle under the influence of distant solitons-particles.

\subsection{Newtonian Potential}
\label{newpotent}

The geodesic line Equation~(\ref{564676361}) in zero speed approximation contains the component \fZ{\metrEff_{00}} only.
Let us write it in the following form:
\begin{equation}
\label{EffffMetr}
\mphantom{
\metrEff_{00} {}={} \pm \left(1 {}-{} 2\,\Nvarphi\right)
\;,\quad
\Nvarphi \eqdef -2\pi\,\xxx^2
\tilde{\cE}
}
\end{equation}
\noindent
where \fZ{\Nvarphi} is the scalar potential of the gravitational field,
\fZ{\tilde{\cE}} is an averaged energy density for the field of distant solitons.
{Here the averaging is performed over a space-time volume including a localization region of the soliton and a relevant time interval}

{In order} to have the real gravitation in our consideration we  {must} obtain the following expression for the gravitational potential:
\begin{equation}
\label{51924319}
\mphantom{
\Nvarphi {}={} -\Ngamma\,\dfrac{\tilde{\mass}}{R}
\;,
}
\end{equation}
where \fZ{\Ngamma} is the gravitational constant, \fZ{\tilde{\mass}} is a mass for agglomeration of distant solitons-particles,
\fZ{R} is a distance from the energy center of this agglomeration.

Thus we must obtain the following asymptotic form for the averaged energy density {as \fZ{R\to\infty}}:
\begin{equation}
\label{831258131}
\mphantom{
\tilde{\cE} \sim \frac{1}{R}
\;.
}
\end{equation}

\subsection{{The} Role of Wave Background}
\label{rolewavb}

Let us elucidate how we can obtain the necessary asymptotic form for the averaged energy density (\ref{831258131}).

{If we consider only static parts of charged solitons-particles (\fZ{\ffun\sim r^{-1}}) then the appropriate energy density
will decrease apparently as \fZ{r^{-4}}.}

{Taking into consideration the oscillating parts of interacting solitons-particles with decreasing amplitude as \fZ{r^{-1}} (as, for example, for the weak soliton (\ref{413418791})) gives the decrease of the
averag{ed} energy density as \fZ{r^{-2}}.}

{%
To obtain the asymptotic \fZ{r^{-1}} for the potential \fZ{\Nvarphi} we must also take into account a wave background with almost constant amplitude in space.
This wave background must undoubtedly exist in the space where there is a bulk of oscillating solitons-particles.}

\def\bFem{\ffun}

 {Then we must} represent the field  {\fZ{\tilde{\bFem}} generating the effective metric} in the form of {the} following sum:
\begin{equation}
\label{513872871}
\mphantom{
\tilde{\bFem} = \tilde{\bFem}^{\circ}+\tilde{\bFem}^{\sim}
\;,
}
\end{equation}
where \fZ{\tilde{\bFem}^{\circ}} is the field of distant solitons containing the {fast} oscillating part with amplitude decreasing as
\fZ{r^{-1}} (just as in the weak soliton (\ref{413418791})),
\fZ{\tilde{\bFem}^{\sim}} is  {the} wave background field with almost constant amplitude.

The substitution of the field (\ref{513872871}) to the energy density gives the terms which are proportional to \fZ{r^{-1}}.
The following averaging of energy density can give the appropriate term in expression for the potential \fZ{\Nvarphi} which is proportional to \fZ{r^{-1}}.
{This expectation seems reasonable because the interaction between the fields \fZ{\tilde{\bFem}^{\circ}} and \fZ{\tilde{\bFem}^{\sim}},
{due to the}
nonlinearity of the model, can
{lead to a certain}
synchronization {of the phases} for these {fast} oscillating fields.
{In this case, the necessary asymptotic term \fZ{r^{-1}} will be kept in the averaged energy density \fZ{\tilde{\cE}} as a result of the averaging of the products
\fZ{\tilde{\bFem}^{\circ}_{0}\,\tilde{\bFem}^{\sim}_{0}}
and \fZ{\tilde{\bFem}^{\circ}_{i}\,\tilde{\bFem}^{\sim}_{i}}.}
}

Thus the gravitational constant is defined by two factors
\begin{equation}
\label{509465151}
\mphantom{
\Ngamma \sim \st{\circ}{\Ngamma}\,\tilde{\Ngamma}
\;.
}
\end{equation}
\noindent
where \fZ{\st{\circ}{\Ngamma}} is a  {constant of proportionality between} an amplitude of the
{fast} oscillating field of distant {agglomeration of} solitons{-particles and its total mass \fZ{\tilde{\mass}}},
\fZ{\tilde{\Ngamma}} defines  {the} amplitude of the {wave} background  {in the considered space-time {region}}.

\section{Practical Applications}
\label{quantbeh}

\subsection{Possible Explanation for the Dark Matter Effect}
\label{posexpldm}

We assume
that the amplitude of the {wave} background
{(the factor \fZ{\tilde{\Ngamma}} in (\ref{509465151}))}
 {may vary slightly} in space.
 In this case we must assume that the gravitational constant  is not {really} constant
{but it can also change slightly}.
Perhaps the observable effect{s} of {so-called} dark matter {\cite{BertoneHooper2018}} can be explained by a {weak} spatial dependence of the gravitational constant~\fZ{\Ngamma (\x)}.

{%
It should also be noted that the concept of gravitational constant is connected with the concept of point mass.
But the effective metric in the induced gravitation theory is defined by the wave background and the field of distant solitons-particles.
The latter
can have various spatial forms
that satisfy the model equation asymptotically.}

{In other words,
for defined space-time configurations of big clusters of massive bodies,
the law of inverse proportional dependence on distance
\fZ{R^{-1}} for
the averaged energy density \fZ{\tilde{\cE}} (\ref{831258131}) can be violated.
Thus the approach under consideration
shows
a similarity with the approach of so called modified Newtonian dynamics
(MOND)~\cite{MilgromM1983}}.

{On the other hand, in the approach under consideration, the gravitational field is induced by the fundamental field of space-time film,
which we consider to be material. However, the oscillating parts of solitons-particles extending far beyond the visible matter do not directly generate the observable radiation. In this sense, we can talk about dark matter. }

The appropriate ideas for explanation of the dark matter effect
require further
study and separate publication.

\subsection{About {the} Possibility of Gravitational Screening}
\label{abposgravscr}

We have  {seen} that {in this approach,}
the existence of {the} long-range gravitational potential \fZ{\sim r^{-1}} is
{ associated with the presence of}
the wave background.

{Then if the wave background is cut off or weakened with the help of some method, then the gravitational interaction will also be weakened.}
In this case we can talk about gravitational screening.

Estimations of possible frequency of the wave background show~\cite{Chernitskii2012be} that the gravitational screening {requires}
a metamaterial {from which a gamma-ray mirror can be made.}

\end{document}